\documentclass[a4paper,12pt,onecolumn,
nofootinbib,
nobibnotes]{iopart}
\usepackage[T1]{fontenc}

\expandafter\let\csname equation*\endcsname\relax

\expandafter\let\csname endequation*\endcsname\relax
\usepackage{bbold}
\usepackage{amssymb}
\usepackage{braket}
\usepackage{mathtools}
\usepackage{graphicx}
\usepackage{xcolor}
\usepackage{tcolorbox}
\usepackage{hyperref}
\hypersetup{citecolor=blue}
\hypersetup{colorlinks=true}

\DeclareMathOperator\rank{rank}
\DeclareMathOperator\diag{diag}
\DeclareMathOperator*\argmin{arg\,min}
\DeclarePairedDelimiter\abs\lvert\rvert
\DeclarePairedDelimiter\rawnorm\lVert\rVert
\newcommand\norm[2]{\rawnorm{#2}_{#1}}
\newcommand{\ketbrad}[1]{\left|{#1}\rangle\!\langle{#1}\right|}
\newcommand{\ketbra}[2]{\left|{#1}\rangle\!\langle{#2}\right|}

\newcommand\new[1]{{\color{black}#1}}

\begin{document}

\title{Comparison of confidence regions for quantum state tomography}

\author{Jessica O.\ de Almeida$^1$, Matthias Kleinmann$^2$, and Gael Sent\'is$^3$}
\address{1. ICFO - Institut de Ci\`encies Fot\`oniques, The Barcelona Institute of Science and Technology, 08860 Castelldefels (Barcelona), Spain.}
\address{2. Naturwissenschaftlich-Technische Fakultät, Universität Siegen, Walter-Flex-Straße 3, 57068 Siegen, Germany}
\address{3. F\'isica Te\'orica: Informaci\'o i Fen\`omens Qu\'antics, Departament de Fisica, Universitat Aut\`onoma de Barcelona, E-08193, Bellaterra (Barcelona), Spain}
\ead{1 jessica.almeida@alumni.icfo.eu}

\begin{abstract}
The quantum state associated to an unknown experimental preparation procedure can be determined by performing quantum state tomography. \new{If the statistical uncertainty in the data dominates over other experimental errors, then a tomographic reconstruction procedure must express this uncertainty.
A rigorous way to accomplish this is via statistical confidence regions in state space}. Naturally, the size of this region decreases when increasing the number of samples, but it also depends critically on the construction method of the region. We compare recent methods for constructing confidence regions as well as a reference method based on a Gaussian approximation. For the comparison, we propose an operational measure with the finding, that there is a significant difference between methods, but which method is preferable can depend on the details of the state preparation scenario.
\end{abstract}

\maketitle

\section{Introduction}

Quantum state tomography is a complete and agnostic procedure to characterize an unknown quantum state from experimental data \cite{Paris2004}. As such, it plays a central role as a quality control subroutine in quantum experiments or quantum technological applications, allowing us to assess the outcome of a state-preparation procedure and verify that a targeted quantum state is indeed being produced.

The outcome of a tomographic procedure is usually a point estimate $\hat\rho$ of the true density matrix $\rho$, along with some sort of error bar. A common approach is to use maximum likelihood estimation \cite{Hradil1997, Rehacek2001} to construct $\hat\rho$, and report the variance of the point estimate obtained by bootstrapping \cite{Efron1994, Home2009} \new{or the width of the likelihood function \cite{Rehacek2008}}. However, this approach accounts only for the statistical fluctuations of the maximum likelihood point estimate, which can turn out to be smaller than its actual bias \cite{Sugiyama2012, Schwemmer2015}. \new{While such error bars do not measure how far off is $\hat\rho$ from the true state $\rho$, they can have use-cases and justification in scenarios where other information on the tomographic experiment is intended. To analyze tomographic data} rigorously in statistical terms, \emph{confidence regions} (CRs) are constructed: these are regions in the space of Hermitian operators that intersect the state space and contain the true state with high probability. Generally speaking, we can regard a confidence region $C$ as a function from the experimental data to the space of Hermitian operators fulfilling
\begin{equation}\label{eq:conf_reg}
\Pr[\,\rho\in C(\mathrm{data})\mid \mathrm{data} \sim\rho\,] \geq 1-\delta
\end{equation}
for every state $\rho$, where $1-\delta$ is the confidence level.

There exist diverse methods to construct CRs from tomographic data in the literature \cite{Blume-Kohout2010, Blume-Kohout2012, Christandl2012, Shang2013, Sugiyama2013,  Faist2016, Wang2018, Guta2020, Franca2021, Araujo2022}. Methods may differ on the theoretical grounds they are built on, in turn leading to differences in their tightness, speed of convergence, and computational cost. However, these features are not always conveyed in a way that allows for a direct comparison between different methods, particularly in the practically relevant regime of finite data. As a result, the natural question of which CR should one choose to report the results of a tomographic experiment has, to this day, no straightforward answer. The main objective of this work is to close this gap by developing an operational benchmark under which methods to build CRs can be fairly contrasted.

Generically, for the same dataset and confidence level, the smaller a CR is, the better. To illustrate why this simple criterion is problematic to evaluate in practice, let us have a first look at two recent results in Refs.~\cite{Wang2018, Guta2020}, which we will further analyze in this work. A usual way of expressing Eq.~\eqref{eq:conf_reg} (see, for example, Refs.~\cite{Sugiyama2013, Guta2020}) is
\begin{equation}\label{eq:conf_reg_2}
    \Pr[\,\norm{\star}{\hat\rho-\rho} \leq \epsilon\,] \geq f(\epsilon,d,N),
\end{equation}
where $d$ is the dimension of the quantum system, $N$ is the number of tomographic experiments, and the form of $f$ varies with the norm $\star$ (usual examples are trace, Hilbert-Schmidt, or operator norm) and with the estimator used for $\hat\rho$. In terms of speed of convergence with the number of samples taken, Guţă \emph{et al.}\ \cite{Guta2020} report an almost optimal scaling of $f$ with respect to $N$, $d$, but fixed $\epsilon$ (showing an improvement over Ref.~\cite{Sugiyama2013} for the same norms). In contrast, Wang \emph{et al.}\ \cite{Wang2018} provide an explicit construction of the function $C$ as a polytope, and the convergence of the region with $N$ is implicit, preventing a direct comparison to Eq.~\eqref{eq:conf_reg_2} based only on the region size. For finite $N$, neither of these works consider the tightness of the CRs quantitatively. Then the question arises, how can we judge whether one of these CRs is smaller than the other in a given experiment? First, the regions have incomparable shapes (a sphere in a norm and a polytope), hence a parametric comparison is not possible. One could look at volumes to disregard shape, but the relevant volume to consider would be that of the intersection of the CRs with the state space, hence we would require numerical estimates (for example, by Monte Carlo sampling) that are not scalable beyond relatively small systems and will be in any case dependent on the chosen integration measure. Finally, we would encounter the same problems when comparing CRs of the form in Eq.~\eqref{eq:conf_reg_2} if these are expressed in different norms. In this case, we could use inequalities between norms to transform one into another, but we would always be underestimating or overestimating one of the regions.

In this work we propose to assess the power of a CR by how well can it distinguish a pair of states. More precisely, we simulate a tomography experiment on two given target states, and find, for a given confidence level, the minimum number of repetitions necessary for their corresponding CRs (built with the same method) to stop intersecting, which ensures that the two states can be distinguished with the chosen confidence.
\new{Intuitively, smaller, more powerful CRs should generally lead to fewer repetitions required for distinguishing two given states, thus allowing us to correlate the statistical power of a CR to a single number}.
While, clearly, this is not the most efficient test to discriminate the two states, as a test it has several desirable features: It has an operational meaning, it is agnostic to the specifics of a CR, and it is computationally efficient. 
\new{The number output by our distinguishability test will certainly depend not only in the confidence level but also on the pair of states chosen; this can be considered to be a feature, since one may tailor the test to assess the power of CRs along a particular direction in state space.}
We use this test to carry out a comparative analysis of recent CRs in paradigmatic scenarios involving pairs of states of up to 4 qubits, and we focus on performing local Pauli-basis measurements as our tomographically complete set of measurements for its practical relevance and widespread use\new{; we also analyse measurements of local symmetric informationally complete positive operator-valued measures (local SIC-POVMs), see~\ref{app:SICpovm}.} Our results show that there is no single CR that outperforms the others in all cases. The performance of the CRs varies not only with the dimension of the system, but also with the pair of states selected for the distinguishability test. An example of this is the finding that while a CR may allow to distinguish a pure state from the completely mixed state with the fewest number of samples, it underperforms when distinguishing two pure states that are very close to each other. This notwithstanding, our analysis allows us to make some general recommendations.

We apply our comparative analysis to the recent results in Ref.~\cite{Guta2020} and Ref.~\cite{Wang2018}. In addition, we provide a method to build CRs  based on a Gaussian approximation, which will serve us as a reference. While there are other works on CRs in the literature, the rationale behind focusing on these ones is the following.
First, we concentrate on unconditional tomography, leaving aside Bayesian methods that require assuming a prior distribution over the possible states (and result in \emph{credibility regions} instead of CRs) \cite{Blume-Kohout2010, Shang2013, Araujo2022}, methods that assume rank-deficient states like compressed sensing \cite{Gross2010}, permutationally invariant states \cite{Moroder2012b}, or matrix product states \cite{Cramer2010a}, and heuristic methods like self-guided tomography \cite{Ferrie2014} or CRs based on likelihood ratios \cite{Blume-Kohout2012}. Second, we select those works which have not yet been clearly superseded (for example, Ref.~\cite{Guta2020} improves on Ref.~\cite{Sugiyama2013}, and Ref.~\cite{Wang2018} on Ref.~\cite{Christandl2012}). Finally, it is worth mentioning with Ref.~\cite{Acharya2019} another recent comparative study on quantum state tomography, which evaluates a different aspect in tomography schemes, namely the typical error (distance between the estimate and the true state) of various point estimators.

The paper is structured as follows. In Section~\ref{sec:crs} we summarize the CRs considered in our study. Section~\ref{sec:methods} details the methodology used for comparing between CRs in a series of simulated tomography experiments, and includes the results of our testing. We end the paper with a short discussion in Section~\ref{sec:discussion}. In the Appendix, we include details on the derivation of the Gaussian CRs as well as technical considerations about the simulations.

\section{Methods for building confidence regions}\label{sec:crs}

Here we summarize the results that we use for our comparative analysis. A common feature of all selected methods is that the CRs are built for the free least-squares estimator for $\hat\rho$, see~\ref{app:adaptations}. Importantly, this estimator is unbiased, although due to finite statistics in experiments it generally produces a non-physical density matrix. This is not problematic, because the associated CR will almost certainly have some overlap with the state space. \new{In fact, if the confidence region does not have an overlap with the state space, this is a clear sign of a misaligned experiment, see Ref.~\cite{Moroder2012b} for an analysis of this point. Hence, even if the point estimate is unphysical, the intersection of the CR with the state space is the physical confidence region, see also Ref.~\cite{Suess2016}.} CRs for physical (hence biased) estimators exist, for example, for \emph{projected} least-squares \cite{Guta2020} or other variants \cite{Acharya2019}, but they are derived from a CR for the free least squares estimator and thus do not improve over them. Hence, to treat all methods on equal footing, in this paper we restrict ourselves to free least squares. For easy referencing, we label the CRs used in our analysis as A, B, C$_1$, and C$_2$.

\medskip

The first result we review is a non-asymptotic concentration bound for the operator-norm error of the free least-squares estimator. In Ref.~\cite{Guta2020}, Guţǎ \emph{et al.}\ show that the free least-squares estimator obeys
\begin{tcolorbox}
{\bf Confidence region A}
\begin{equation}
\Pr[\, \norm{\infty}{\hat \rho-\rho}\ge \epsilon\,]
\le R_d\left(\epsilon\sqrt{N}\right),
\quad\text{for}\quad 0\le \epsilon\le 1
\end{equation}
\end{tcolorbox}
\noindent where $\norm{\infty}X$ is the operator norm of $X$, $N$ is the total number of state preparations, and $d$ is the dimension of the underlying Hilbert space. Hence, the CR is of the form in Eq.~\eqref{eq:conf_reg}. One has
\begin{equation}
\log R_d(\xi)=-\tfrac38\xi^2 g(d)+\log d,
\end{equation}
where the factor $g(d)$ takes different forms depending on the chosen measurement scheme. For multiqubit states and local Pauli-basis measurements, \new{one has $g(d)= 3^{-\log_2 d}$, for SIC-POVM measurements, $g(d)=2d$, but no expression for $g(d)$ has been provided for the case of local SIC-POVM measurements of two or more qubits}.

\medskip

Another method by Wang \emph{et al.}\ \cite{Wang2018} constructs a polytope-shaped region $\Gamma(\vec f)$ as a function of the empirical frequencies $\vec{f}=(f_1,\dotsc,f_M)$ of a single \textit{positive operator-valued measure} (POVM), that is, positive semidefinite operators $\{E_i\}_{i=1}^M$ obeying $\sum_i E_i = \mathbb{1}$, applied on the state $\rho$, where $f_i=n_i/N$, $n_i$ is the number of occurrences of outcome $i$, and $N$ the total number of state preparations. The region $\Gamma(\vec f)$ obeys
\begin{tcolorbox}
{\bf Confidence region B}
\begin{equation}\label{eq:bound_renner}
    \Pr[\,\rho\notin \Gamma(\vec{f})\,]\le \delta.
\end{equation}
\end{tcolorbox}
\noindent Each facet of the polytope is determined by the half-space
\begin{equation}\label{eq:polytope_states}
    \Gamma_i(f_i)=\set{\rho | \tr(E_i \rho)\leq f_i+\epsilon_i },
\end{equation}
where $\epsilon_i$ is the positive root of
$D(f_i\Vert f_i+\epsilon_i)=-\log(\delta_i)/N$,
with $\sum_{i=1}^M \delta_i = \delta$, and $D(x\Vert y) = x \log (x/y) + (1-x)\log[(1-x)/(1-y)]$ is the relative binary entropy. Then, the CR is given by the intersection of all half-spaces, that is, $\Gamma(\vec f)=\bigcap_i \Gamma(f_i)$. Because each half-space corresponds to an independent confidence interval for the frequency $f_i$, we are free to choose each individual confidence level by fixing $\delta_i$ as long as they add up to the total confidence parameter $\delta$. This method thus allows us to shrink the CR along a preferred measurement direction at the expense of enlarging it in other ones, which adds versatility. For our comparative analysis, we make the uniform choice $\delta_i=\delta/M$.

\medskip

We also provide bounds based on a Gaussian approximation. Here we make the approximation that the empirical frequencies $\vec f$ of the measurement outcomes are Gaussian distributed with mean $\vec p(\rho)$ and a nonsingular covariance matrix $\Sigma(\rho)$. Then we obtain the Gaussian CR defined by
\begin{tcolorbox}
{\bf Confidence region C$_1$}
\begin{equation}
  \Pr[\, \norm2{\Sigma(\rho)^{-\frac12}(\vec f-\vec p(\rho))} > \epsilon\,]=P_k(\epsilon^2),
\end{equation}
\end{tcolorbox}
\noindent where $P_k(x)$ is the survival function of the central $\chi^2$-distribution with $k=\rank(\Sigma)$ degrees of freedom, see~\ref{app:derivation_bound_2} for details. Note that $\Sigma(\rho)$ here scales with the number of samples. This CR is exact, but comes with the disadvantage that the region depends nonlinearly on $\rho$ through $\Sigma$. We can improve on this at the cost of getting a more relaxed bound, see~\ref{app:derivation_bound_2},
\begin{tcolorbox}
{\bf Confidence region C$_2$}
\begin{equation}\label{eq:bound22}
  \Pr[\, \norm2{\vec p(\hat \rho)-\vec p(\rho)}
         >\epsilon \,]\le P_{d^2-1}(\sigma^{-2}_\mathrm{max}\epsilon^2),
\end{equation}
\end{tcolorbox}
\noindent with $\sigma^2_\mathrm{max}$ the largest eigenvalue of $\Sigma(\rho)$ across all states.  Note that the resulting CR forms an ellipsoid in the state space. For multinomial distributions, we have $\sigma^2_\mathrm{max}=1/(2n)$, where $n$ is the number of samples per measurement setting. For example, for local Pauli-basis measurements on a $q$-qubit state, we have $3^q$ distinct measurement settings, $N=3^q n$ (see~\ref{app:derivation_bound_2}).

\section{Methodology for comparing CRs and results}\label{sec:methods}

We propose two universal tests to benchmark CRs irrespective of their specifications. The first test consists in computing the $(1-\delta)$-quantile of the distribution of the ratio
\begin{equation}\label{eq:quantile}
r=\frac{\norm{\star}{\dots}}{\epsilon(\delta)}
\end{equation}
over several repetitions of the same tomography experiment with confidence level $(1-\delta)$. The terms in $r$ depend on the chosen CR, as they appear in $\Pr[\,\norm{\star}{\dots} > \epsilon(\delta)\,] \leq \delta$ and note that the norm can be defined either in probability space or in the state space. This test quantifies the tightness of a CR in a scale-independent way. On the one hand, a value 100\% for the $(1-\delta)$-quantile of $r$ indicates that the region is exact, inasmuch as the probability of obtaining $\hat\rho$ at a distance no more than $\epsilon$ from $\rho$ coincides with the confidence level, that is, the inequality in Eq.~\eqref{eq:conf_reg_2} is saturated. On the other hand, a small value of $r$ tells us that the CR is unnecessarily large, since in $(1-\delta)\times 100\%$ of the experiments $\hat\rho$ is actually much closer to $\rho$ than the size of the region $\epsilon$.
\begin{table}
\centering
\begin{tabular}{l|r|r|r|r}
$(1-\delta)$-quantile&1 qubit&2 qubits&3 qubits&4 qubits \\\hline
CR A      &  38\% $\pm$ 2\% &  \new{41\% $\pm$ 3\%} &  \new{36\% $\pm$ 1\%} &  \new{32\% $\pm$ 1\%}\\
CR C$_1$  & 100\% $\pm$ 2\% & 100\% $\pm$ 1\% & 100\% $\pm$ 3\% & 100\% $\pm$ 2\%\\
CR C$_2$  &  61\% $\pm$ 3\% &  58\% $\pm$ 6\% &  45\% $\pm$ 3\% &  \new{33\% $\pm$ 1\%}\\
\end{tabular}
\caption{Empirical $(1-\delta)$-quantile of the distribution of the ratio $r$ defined in Eq.~\eqref{eq:quantile} for different CRs and pure Haar-random states \new{for tomography based on Pauli-basis measurements}. We used $N=60\,000$ samples on each tomography simulation, and evaluated the $(1-\delta)$-quantile from 10$\,$000 samples of $r$. A perfect region reaches 100\% while a lower value shows that the CR is too large. The provided errors show the range of the values attained for 50 different random states.}
\label{tab:quantiles}
\end{table}

We run simulations of tomography experiments on Haar-random pure states of up to 4 qubits. We simulate using $N=60\,000$ samples and local Pauli-basis measurements, see~\ref{app:localpaulis}, and evaluate $r$ for a confidence level of $1-\delta=99\%$. For Confidence Region~A, Confidence Region~C$_1$ and Confidence Region~C$_2$, and for each state considered, we determine the empirical $(1-\delta)$-quantile from $10\,000$ samples of the ratio $r$; the results are summarized in Table~\ref{tab:quantiles}. Note that, since Confidence Region~B is not norm-based, we do not run this test on it\new{; we show here the results for Pauli-basis measurements, in~\ref{app:SICpovm} we provide the quantiles for local SIC-POVMs, and Confidence Regions~C$_2$ and A (the latter only for 1-qubit cases)}.
We observe that Confidence Region~C$_1$ is an exact region, which supports the validity of the Gaussian sampling assumption in our numerical experiments. For Confidence Region~A and C$_2$ we see that \new{the latter} is tighter for the 1-qubit, 2-qubit and 3-qubit cases and then becomes looser for states of 4 qubits, whereas the tightness of Confidence Region~A is more stable with an increasing number of qubits. As we see next, these results correlate well with those of our second test, even though they quantify different features of the CRs.

\begin{figure}
\centering
\includegraphics[scale=0.5,keepaspectratio]{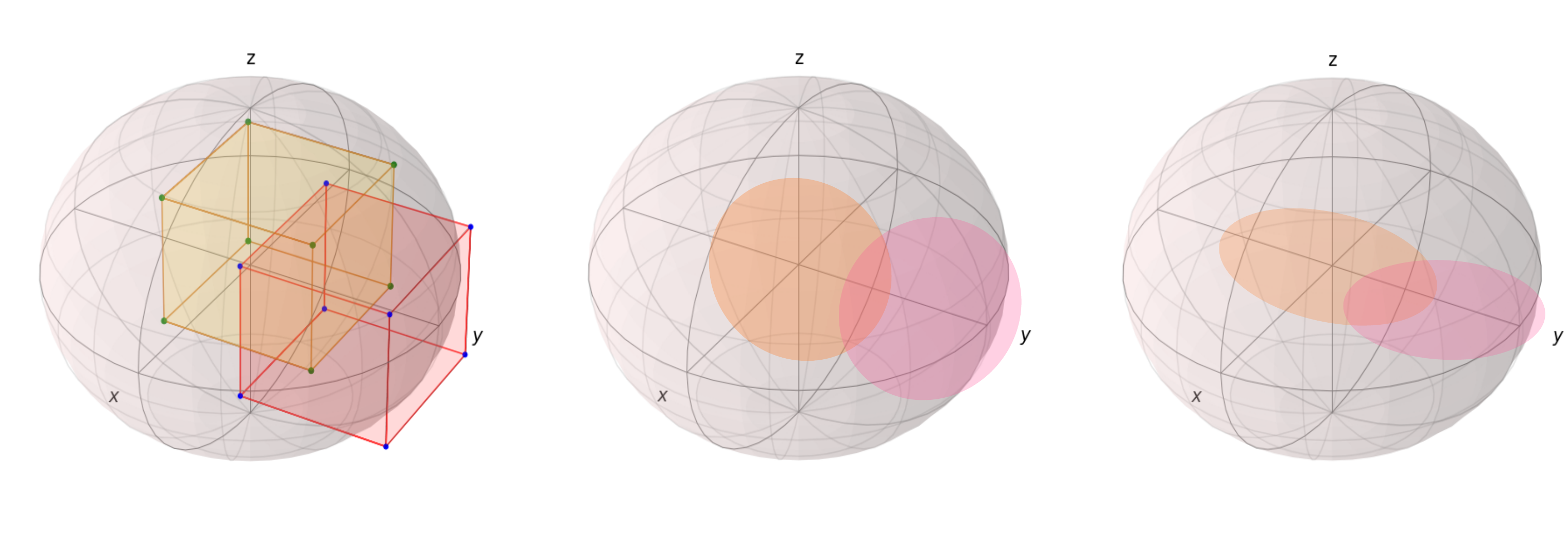}
\caption{Intersecting CRs of different shapes corresponding to tomography experiments performed for two one-qubit states. CRs of type B are polytopes (left); CRs of type A are spheres, although they need not be for higher-dimensional systems (center); and CRs of type C$_2$ are generally ellipsoids (right).}
\label{fig:intersections_scheme}
\end{figure}

The second test we devise is aimed at quantifying the distinguishability power of CRs. Choosing two distinct target states over which we perform tomography, we ask the question of how many samples $N$ are required so that their respective CRs do not contain a common state. At this point we can distinguish the two states with confidence $1-\delta$ based on the CRs, see Figure~\ref{fig:intersections_scheme}. In this way we can attach a number to a CR and a given pair of states which is indirectly related to the size of the region. This test thus allows us to rank CRs according to an operationally well defined task and independently of their specifications.

We run this test with Confidence Region~A, B, and C$_2$ for two types of pairs of states of up to 4 qubits: pairs comprised by a pure state and the completely mixed state, and pairs of two pure states relatively close to each other, with fidelity $\abs{\braket{\phi_1|\phi_2}}^2= 0.63$. The specific states are detailed in~\ref{app:states}, and we fix the confidence level to $1-\delta=90\%$.
Importantly, the value of $N$ at which the CRs stop intersecting is a stochastic variable, 
since it directly depends on the estimate $\hat\rho$ where the regions are centered and thus on the data obtained in each tomography experiment. For increasing values of $N$, we look at the frequency of finding an intersection over many repetitions (3000 for $q=1,2$; 1000 for $q=3$ and 500 for $q=4$). We also note that testing whether two regions share a common state can be computed efficiently via a convex optimization program, see~\ref{app:sdp_intersections}. This is the case with the exception of Confidence Region~C$_1$ since this region depends nonlinearly on the true state through $\Sigma(\rho)^{-\frac12}$, and for this reason we do not include it in this test.

\begin{figure}
\centering
\includegraphics[scale=0.45]{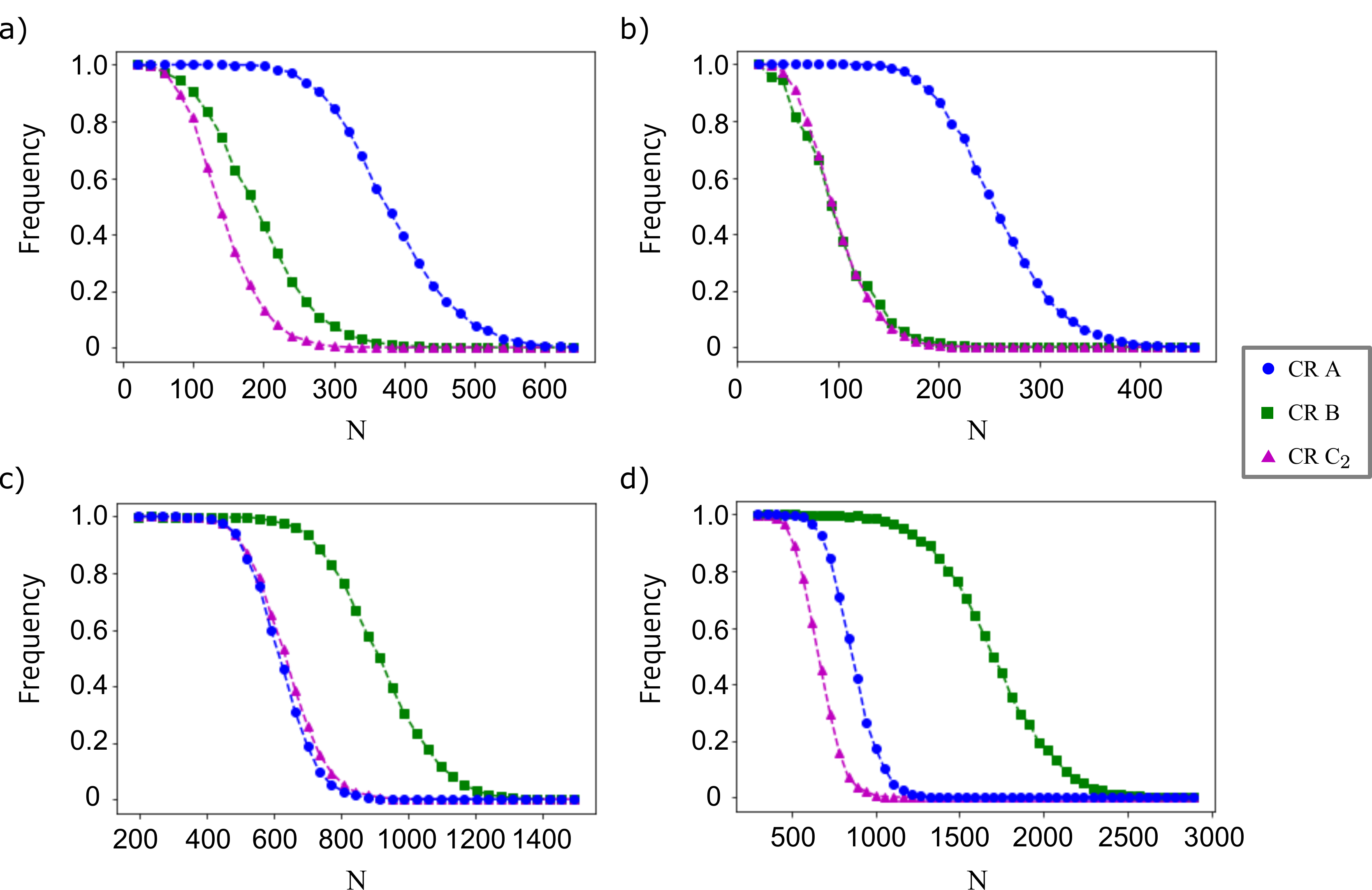}
\caption{Relative frequency of finding a state in the intersection between the CRs of
a) (1 qubit) a pure state and the completely mixed state,
b) (1 qubit) two pure states with fidelity 0.63,
c) (2 qubits) a rotated Bell state and the completely mixed state, and
d) (2 qubits) two rotated Bell states with fidelity 0.63.
See also~\ref{app:states} for details. \new{The tomography is here performed with Pauli-basis measurements. For the corresponding analysis using local SIC-POVM measurements, see~\ref{app:SICpovm}.}}
\label{fig:1and2qubits}
\end{figure}

In Figure~\ref{fig:1and2qubits} we show the frequency that the CRs of the two target states contain a common state, for each of the three CRs considered \new{and for tomography using Pauli-basis measurements}. Each plot corresponds to a different pair of target states of one and two qubits. For the case of one-qubit states, Confidence Region~C$_2$ performs best in this test, closely followed by Confidence Region~B, regardless of whether we try to distinguish a pure state versus the completely mixed state or two pure states close to each other, whereas Confidence Region~A requires roughly a factor of two more state preparations. For two-qubit states we observe an interesting phenomenon: Confidence Region~A distinguishes a Bell state from the identity with the least number of state preparations and Confidence Region~C$_2$ has almost the same performance, but it becomes notably more inefficient when distinguishing between a Bell state and the same state with a slight local rotation applied (with fidelity 0.63), for which case Confidence Region~C$_2$ is again the best one. An important remark is that the states selected for the test are not aligned with any measurement direction, in order to represent a generic case of state tomography. Particularly for one-qubit states, if the state would point in the $x,y$ or $z$ direction, we find that Confidence Region~B slightly outperforms Confidence Region~C$_2$. This is due to the facets of the polytope $\Gamma(\vec f)$ being perpendicular to the measurement directions by design.

\begin{figure}
\centering
\includegraphics[scale=0.45]{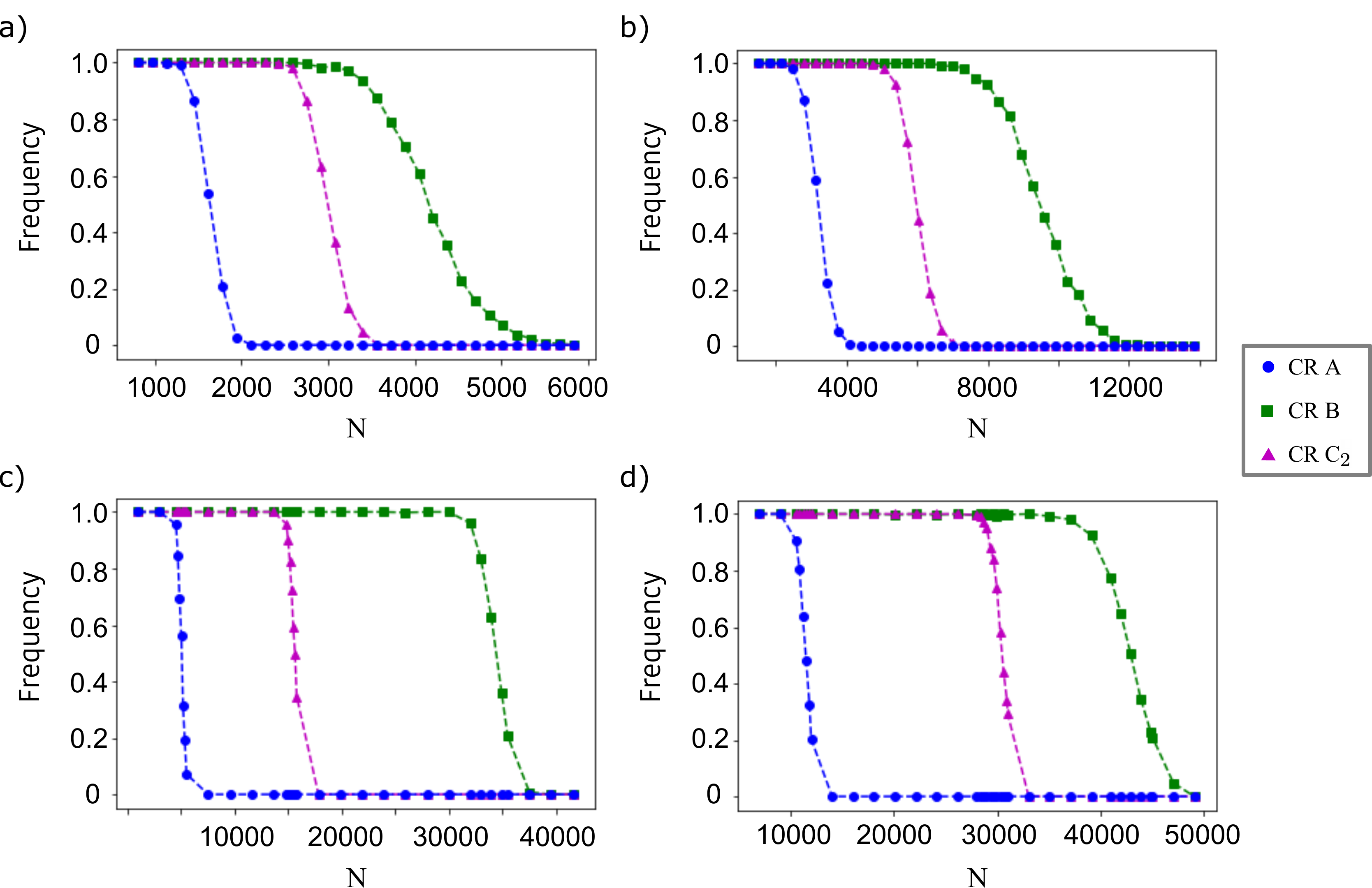}
\caption{Relative frequency of finding a state in the intersection between the CRs of
a) (3 qubits) a rotated Greenberger--Horne--Zeilinger (GHZ) state  and the completely mixed state,
b) (3 qubits) two rotated GHZ states with fidelity 0.63,
c) (4 qubits) a rotated entangled state and the completely mixed state, and
d) (4 qubits) two rotated entangled states with fidelity 0.63.
See also~\ref{app:states} for details.}
\label{fig:3and4qubits}
\end{figure}

We show the results of the intersection test for pairs of 3-qubit and 4-qubit states in Figure~\ref{fig:3and4qubits}\new{, again for Pauli-basis measurements}. Here we observe a clear advantage of Confidence Region~A in all cases. This is consistent with the results of Ref.~\cite{Guta2020}. Indeed, the derivation of Confidence Region~A is based on the observation that $\hat\rho$ can be seen as a sum of $N$ random matrices and the subsequent application of a matrix concentration inequality, which becomes sharper as $N$ increases (which is necessary for a larger number of qubits). Also, the ranking between the analyzed CRs is maintained throughout the studied 3-qubit and 4-qubit cases, with Confidence Region~C$_2$ in a second position and Confidence Region~B requiring the largest number of state preparations.

\new{We also apply the same analysis as in Figure~\ref{fig:1and2qubits} and Figure~\ref{fig:3and4qubits} for measurements of local SIC-POVMs, see~\ref{app:SICpovm}.
The results for one qubit are essentially comparable to Pauli basis measurements, although Confidence Region~B performs slightly better. When considering two or more qubits, we find that Confidence Region~C$_2$ becomes slightly worse while Confidence Region~B becomes worse in distinguishing pure states from the completely mixed state, but gets significantly better in distinguishing two pure states. Generally, Confidence Region~B outperforms C$_2$ on the usage of SIC-POVMs for 3 or more qubit systems. Note that for two and more qubits we cannot apply the analysis to Confidence Region~A, since the factor $g(d)$ is not known for local SIC-POVMs.}

\section{Discussion}\label{sec:discussion}

We have proposed and tested two methods to benchmark confidence regions for quantum state tomography. One of them, based on the distinguishability power of a given confidence region construction, is operationally defined and applicable to any such construction regardless of its shape, specifications, and whether the region is defined in probability space or in state space. In addition, we proposed a second method which determines how tight a given region is. These new metrics allow us to compare existing (and future) methods to build confidence regions that were until now not directly comparable. Thus, we provide a robust tool to help in deciding which confidence region to report in any state tomography experiment.

Our numerical experiments show that this choice should depend on the dimension of the system at hand and the goal for which tomography is used. We see this variability as a reflection of the diverse techniques used to derive confidence regions. \new{The method by Guţǎ \emph{et al.}\ \cite{Guta2020} appears to perform particularly well for large systems, while in special cases involving 1-qubit and 2-qubit states, the method of Wang \emph{et al.}\ \cite{Wang2018} or a Gaussian approximation as in Confidence Region~C$_2$ can be better suited.}
In addition, Confidence Region~C$_2$ has also been found to be useful in deriving optimal tests for detecting properties of the measured state directly from the tomographic data, without going through the intermediate step of computing a point estimate of the density matrix. An example of this is a proposal for the experimental detection of bound entanglement \cite{Sentis2018a}. 

\new{We argued that the three confidence regions A \cite{Guta2020}, B \cite{Wang2018}, and C$_2$ are superior to previous regions and should be taken in consideration when evaluating tomographic data. Among those regions, there are differences that can make one preferable over the other. From our experience (with simulated data), one can generally recommend to consider Confidence Region~A first, if the measurement scheme in the experiment belongs to the cases for which the factor $g(d)$ is known. As a fallback, one can consider Confidence Region~C$_2$; improvements on this bound may be subject to further research. The Confidence Region B yields a mixed experience, but can be very good, in particular for measurements of local SIC POVMs. This region can be more difficult to handle and to grasp intuitively, because it yields a polytope with many facets in a high-dimensional space.}

\new{We analyzed recent unconditional confidence regions for full quantum state tomography and we found that the considered regions perform already reasonably well in the light of our quantile tests and are considerably easy to apply.} A point estimate of the unknown state obtained by linear inversion equipped with a confidence region is computationally inexpensive and should suffice for any purpose where tomography is needed. As for which confidence region is best to use, as we have seen, it depends. \new{However,} there may still be room for improvement, \new{since, for example, for the method of Guţǎ \emph{et al.}\ we observe the $(1-\delta)$-quantile is only around 40\%}. This means there could be an even tighter construction, although this seemingly suboptimal slackness does not have a great effect on our intersection tests.

\section*{Acknowledgments}
We thank Carlos de Gois, Richard Kueng, Maciej Lewenstein, and Michael Skotiniotis for the helpful discussions. This work was supported by ERC (Advanced Grant NOQIA and Consolidator Grant 683107/TempoQ); Ministerio de Ciencia y Innovation Agencia Estatal de Investigaciones
(PGC2018-097027-B-I00/10.13039/501100011033, CEX2019-000910-S/10.13039/ 501100011033, Plan National FIDEUA PID2019-106901GB-I00, Plan National STAMEENA PID2022-139099NB-I00 project funded by MCIN/AEI/10.13039/501100011033 and by the “European Union NextGenerationEU/PRTR”
(PRTR-C17.I1), FPI, QUANTERA MAQS PCI2019-111828-2, QUANTERA DYNAMITE PCI2022-132919, Proyectos de I+D+I ‘Retos Colaboración’ QUSPIN RTC2019-007196-7, PID2019-107609GB-I00); MICIIN Ministry of Economic Affairs and Digital Transformation of the Spanish Government through the
QUANTUM ENIA project call Quantum Spain project, and by the European Union through the Recovery, Transformation, and Resilience Plan - with funding from European Union NextGenerationEU(PRTR-C17.I1) within the framework of the Digital Spain 2026 Agenda and by Generalitat de Catalunya; Fundació Cellex; Fundació Mir-Puig; Generalitat de Catalunya (European Social Fund FEDER and CERCA program, AGAUR Grant No. 2021 SGR 01452, QuantumCAT U16-011424, QuantumCAT 001-P-001644 co-funded by ERDF Operational Program of Catalonia 2014-2020); Barcelona Supercomputing Center MareNostrum (FI-2023-1-0013); EU Horizon 2020 FET-OPEN OPTOlogic (Grant No. 899794); EU Quantum Flagship (PASQuanS2.1, 101113690); EU Horizon 2020 FET-OPEN OPTOlogic (Grant No 899794); EU Horizon Europe Program (Grant Agreement 101080086 – NeQST), National Science Centre, Poland (Symfonia Grant No. 2016/20/W/ST4/00314); ICFO Internal ‘QuantumGaudi’ project; European Union’s Horizon 2020 research and innovation program under the Marie-Skłodowska-Curie Grant Agreement Nos. 101029393 (STREDCH) and 847648; ‘La Caixa’ Junior Leaders fellowships ID100010434: CF/BQ/PR23/11980043; Deutsche Forschungsgemeinschaft (DFG, German Research Foundation, Project Numbers 447948357 and 440958198); Sino-German Center for Research Promotion
(Project M-0294); German Ministry of Education and Research (Project QuKuK, BMBF Grant No. 16KIS1618K). Views and opinions expressed in this work are, however, those of the author(s) only and do not necessarily reflect those of the European Union, European Climate, Infrastructure and Environment Executive Agency (CINEA), nor any other granting authority. Neither the European Union nor any granting authority can be held responsible for them.

\section*{References}
\bibliographystyle{iopart-num}
\bibliography{new}

\appendix

\section{Confidence region for normal-distributed tomographic data}\label{app:derivation_bound_2}

Here we provide the proofs for Confidence Region~C$_1$ and Confidence Region~C$_2$. For that, we write $\vec f$ for the collection of the empirical frequencies and assume that $\vec f$ is normal distributed with mean $\vec p(\rho)$ and covariance matrix $\Sigma(\rho)$, where $\rho$ is the state on which the tomography is performed. Statistically dependent entries are eliminated from $\vec p$ and $\vec f$---for example for a multinomial distribution, $\sum_i p_i=1$ and hence we remove one entry from $\vec p$. Then $\Sigma(\rho)$ is nondegenerate, except for a subset of states which has measure zero, and hence is irrelevant.

We observe that
\begin{equation}
    \vec \Delta_\rho = \Sigma(\rho)^{-\frac12} [\,\vec p(\rho)-\vec f\,]
\end{equation}
is normal distributed with mean $\vec 0$ and covariance matrix $\mathbb{1}$. Consequently $\norm2{\vec \Delta_\rho}^2$ is distributed according to the central $\chi^2(k)$-distribution with $k=\rank(\Sigma)$ degrees of freedom and survival function $\Pr[\,\norm2{\vec \Delta_\rho}^2 > y\,]=P_k(y)$. This already yields Confidence Region~C$_1$ via
\begin{equation}
  \Pr[\, \norm2{\vec \Delta_\rho} > \epsilon\,] =  P_k(\epsilon^2).
\end{equation}
We mention that for a multinomial distribution and $n$ samples one has
\begin{equation}
 (\vec \Delta_\rho)^2= n \sum_i \frac{[\,p_i(\rho)-f_i\,]^2}{p_i(\rho)},
\end{equation}
where the sum includes the element that had been omitted from $\vec p$ and $\vec f$ in order to make $\Sigma(\rho)$ nondegenerate.

For Confidence Region~C$_2$ we consider $M(\rho-\rho')=\vec p(\rho)-\vec p(\rho')$ as linear map $M$ on the space of zero-trace Hermitian matrices, which we consider as a real vector space of dimension $d^2-1$. We write $M^+$ for the pseudo-inverse of $M$ and $P$ is a map onto $\mathbb R^{d^2-1}$ with $P=PMM^+$ and $M=P^TPM$. Note that $P$ can can be constructed by orthogonalizing the row vectors of $M^T$. We use now the free least-squares estimate
\begin{equation}\label{eq:ourlsq}
  \hat \rho=\argmin_X\set{\norm2{\vec p(X)-\vec f}^2 | X=X^\dag \text{ and } \tr(X)=1}
\end{equation}
where the result is not required to be positive semidefinite. We have $\hat \rho=M^+\vec f$ and $\rho=M^+\vec p(\rho)$. It follows that $PM(\hat\rho -\rho)=PMM^+[\vec f-\vec p(\rho)]=P[\vec f-\vec p(\rho)]$ has mean $\vec 0$ and covariance $\Sigma'(\rho)=P\Sigma(\rho)P^T$. Note that $\Sigma'(\rho)$ is nonsingular (even if $\Sigma(\rho)$ is degenerate) because $M$ is injective and has dimensions $(d^2-1)\times(d^2-1)$. Therefore we have
\begin{equation}
  \Pr[\, \norm2{\Sigma'(\rho)^{-\frac12}P(\vec p(\hat\rho)-\vec p(\rho))} >\epsilon \,] = P_{d^2-1}(\epsilon^2)
\end{equation}

The main advantage of both confidence regions derived so far is that they do not contain any approximations up to the assumption of a normal distribution. Hence they are optimal in this respect. However, they do not have a simple shape because they depend in a nonlinear way on $\rho$ via $\Sigma(\rho)$. For this reason we replace $\Sigma(\rho)$ by $\sigma_\mathrm{max}^2\mathbb{1}$ with $\sigma^2_\mathrm{max}$ the largest eigenvalue of the covariance matrix over all states $\rho$. Using $P^TP[\vec p(\hat \rho)-\vec p(\rho)]= \vec p(\hat \rho)-\vec p(\rho)$, this gives us Confidence Region~C$_2$ via
\begin{equation}
  \Pr[\, \norm2{\vec p(\hat \rho)-\vec p(\rho)}
         >\epsilon \,]\le P_{d^2-1}(\sigma^{-2}_\mathrm{max}\epsilon^2).
\end{equation}
\new{A simple upper bound for $\sigma^2_\mathrm{max}$ is given by $1/2n$, where $n$ is the number of samples per measurement setting.
In the case of nonprojective measurements, one can use the upper bound $\lambda_\mathrm{max}/n$, where $\lambda_\mathrm{max}$ is the maximal eigenvalue of all measurement effects, for example, for local SIC-POVMs this would be $\sigma^2_\mathrm{max}=2^{-q}/n$. This is the case, because $\Sigma \le \diag(\vec p)/n \le \lambda_\mathrm{max}(\vec p)\mathbb{1} /n$.}
We mention that the spectral decomposition of $\epsilon\, (M^TM)^{-\frac12}$ yields the semiaxes of the confidence ellipsoid in the state space.

We can relax Confidence Region~C$_2$ even further. For that we let $\mu$ be the smallest (nonzero) singular value of $M$. Since $M$ is injective, it follows that $\norm2{M(\rho-\hat \rho)}\ge \mu\norm2{\hat \rho-\rho}$. This gives us an additional CR defined by
\begin{equation}
\Pr[\, \norm2{\rho-\hat \rho}> \epsilon \,]
  \le P_{d^2-1}(\mu^2 \sigma_\mathrm{max}^{-2} \epsilon^2).
\end{equation}
Although this bound has a very similar structure to Confidence Region~A, it cannot be expected to yield a good CR. This can be understood geometrically, because we approximated the previous CR, which gave us an ellipsoid, by simply fitting a sphere around it with the radius of the largest half-axis.

\section{Local Pauli-basis measurements for quantum state tomography}\label{app:localpaulis}

In this appendix we provide specifications for computing the CRs analyzed in the main text when choosing local Pauli-basis measurements as a tomographically complete set of measurements.

Generically, a measurement on a $d$-dimensional quantum state $\rho$ can be described by a POVM, that is, as 
$M$ positive semi-definite matrices $E_1,\dotsc,E_M$, each with dimension $d\times d$, such that $\sum_i E_i=\mathbb{1}$. Each measurement of $\rho$ results in a corresponding outcome and the probability of observing outcome $i=1,\ldots,M$ is given by $p_i=\tr( E_i \rho )$. This probability can be estimated based on the outcome frequency when measuring many copies of $\rho$. The frequency corresponding to outcome $i$ after $n$ repetitions of the measurement $E_1,\dotsc,E_M$ is
\begin{equation}
f_i= \frac{n_i}{n} \qquad \text{for } i=1,\dotsc,M \,,
\end{equation}
and we have $f_i\to p_i$ in the limit $n\to \infty$.

\subsection{Local Pauli-basis measurements}

We now particularize this generic POVM for state tomography with local Pauli-basis measurements. For one qubit, measuring in the three Pauli bases can be viewed as a single six-outcome POVM, whose elements correspond to the eigenstates of each Pauli matrix $\sigma_j$, $j=x,y,z$. Each POVM element has two labels, the direction $j$ and the two possible outcomes along that direction, $+1$ and $-1$. Therefore, we  write $E_{j,\pm}=(\mathbb{1}\pm\sigma_j)/2=\ketbrad{w_{j,\pm}}$. When each measurement $\sigma_j$ is performed $n$ times, the frequency vector corresponding to the six-outcome POVM has components $f_{j,\pm}=n_{j,\pm}/(3n)$.

Generalizing the tomography scheme for $q$ qubits, the dimension of the system is $d=2^q$, and there are $3^q$ Pauli measurement settings $\vec m=(m_1,\dotsc,m_{q})\in \set{x,y,z}^q$, each with $d$ possible outcomes $\vec o = (o_1,\dotsc,o_{q})$, where $o_j\in \set{\pm 1}$. This tomography scheme can also be written as a single POVM, whose elements are now specified by the measurement setting $\vec m$ and the outcome $\vec o$, as
\begin{equation}
 E_{\vec{m},\vec{o}} = 3^{-q}\bigotimes_{j=1}^q \ketbrad{w_{m_j,o_j}}.
\end{equation}
If each measurement setting $\vec m$ is performed $n$ times, we obtain for the single POVM the frequencies $f_{\vec m,\vec o} = n_{\vec m,\vec o}/(3^q \, n)$ where the total number of samples is $N=3^q\,n$.

At this point we note that, formally, there is no difference between measuring each combination of Pauli matrices $\vec m$ separately as $2^q$-outcome projective measurement and considering the single POVM $E_{\vec m,\vec o}$ with $6^q$ outcomes. However, in the finite data regime there is a potential difference when it comes to the covariance matrix of the data. Indeed, when considering a single POVM there is only one frequency that is dependent of the others, by the normalization $\sum_{\vec m,\vec o} f_{\vec m,\vec o}=1$. If, instead, we consider applying the Pauli measurements specified by each $\vec m$ independently and aggregating the data at a later stage, we have now $3^q$ dependent frequencies, since now $\sum_{\vec o} f_{\vec m,\vec o} = 1$. This is relevant when computing the covariance matrix $\Sigma(\rho)$ that appears in Confidence Region~C$_1$: if we do not eliminate frequencies corresponding to the trivial outcomes from $\vec f$, $\Sigma(\rho)$ will not have a well defined inverse. Moreover, as we saw in Appendix~\ref{app:derivation_bound_2}, discarding these outcomes results in a tighter CR. When detailing how to compute Confidence Region~A and Confidence Region~B for local Pauli-basis measurements, we use the single-POVM formulation to follow the original references \cite{Guta2020} and \cite{Wang2018} closer. We switch to the individual measurement settings $\vec m$ when discussing our bound Confidence Region~C$_2$ \eqref{eq:bound22}.

\subsection{Adaptations for the Confidence Regions}\label{app:adaptations}

\paragraph{Generalized Bloch representation.}
It is convenient for our computations to write the density matrix in the generalized Bloch representation. For the tuple of $4^q$ matrices $\vec{\mu}=2^{-\frac q2}(\mathbb{1},\sigma_x,\sigma_y,\sigma_z)^{\otimes q}$, any Hermitian matrix can be written as $X = \vec{x}\cdot\vec{\mu}$ with $\vec x\in \mathbb R^{4^q}$ . Due to the choice of the normalization, we have $\tr(XY)=\vec x\cdot\vec y$ for $X=\vec x\cdot\vec \mu$ and $Y=\vec y\cdot\vec\mu$.

\medskip
\paragraph{Free least-squares estimator.}
The free least-squares estimator in state tomography is the solution to the least-squares problem that seeks to minimize the distance between the frequencies and the probabilities, see also Eq.~\eqref{eq:ourlsq},
\begin{equation}
\hat{\rho}=\argmin_X\Set{ \sum_i \left( f_i- \tr( E_i X )\right)^2 | X=X^\dag \text{ and } \tr(X)=1 }.
\end{equation}
Note that the optimization is performed over all Hermitian matrices with unit trace and hence the resulting estimate $\hat \rho$, in general, will fail to be positive semidefinite. For local Pauli-basis measurements, where $\vec f$ has $6^q$ entries, this problem has an analytical solution \cite{Guta2020}. In our definitions, the frequencies have the form $f_{\vec m,\vec o}= n_{\vec m, \vec o}/(3^q n) $, and the free least-squares estimate is given by
\begin{equation}
\hat{\rho}=\sum_{\vec{m},\vec{o}}f_{\vec{m},\vec{o}}\bigotimes_{j=1}^q \left(3\ketbrad{w_{m_j,o_j}} -\mathbb{1}\right).
\end{equation}
We use this explicit solution when evaluating Confidence Region~A.

\medskip
\paragraph{Confidence Region C$_1$ and C$_2$.}
As indicated in Appendix~\ref{app:localpaulis}, when considering Confidence Region~C$_1$ and Confidence Region~C$_2$, we assume that each measurement setting $\vec m$ is measured independently $n$ times, yielding frequencies $f_{\vec m,\vec o}=n_{\vec m,\vec o}/n$ with $\sum_{\vec o} f_{\vec m,\vec o} = 1$. Therefore, there is one statistically dependent frequency per measurement setting $\vec m$, which we can compute from the remaining frequencies of the same setting. We thus omit the $3^q$ dependent frequencies to avoid having a singular covariance matrix, which reduces the dimension of the data $\vec f$ to $3^q(2^q-1)$ components.

\section{Intersection of confidence regions via convex optimization}\label{app:sdp_intersections}

Given two regions of the types A, B, or C$_2$, we can determine whether they have a non-empty intersection via convex optimization, see, for example, Ref.~\cite{Optimization2004}. In our numerical experiments we use the Splitting Conic Solver SCS \cite{SCS} for convex-cone optimization with the Python programming language.

\medskip
\paragraph{Intersection for Confidence Region~A.}
We have two CRs, $R_1$ and $R_2$, for two tomography experiments on the states $\rho_1$ and $\rho_2$, and corresponding free least-squares estimates $\hat\rho_1$ and $\hat\rho_2$. Confidence Region~A for confidence level $1-\delta$ is given by all states $\rho$ obeying
\begin{equation}
    \norm{\infty}{\rho-\hat \rho_i}\le \epsilon(\delta),
\end{equation}
with $\epsilon(\delta)=N^{-\frac12}R_{2^q}^{-1}(\delta)=\sqrt{3^q \log(2^q/\delta)\,8/(3N)}$ and $N$ the total number of samples. Care has to be taken that $\epsilon\le 1$ holds, which translates for $1-\delta=99\%$ and 1, 2, 3, and 4 qubits to $N\ge 43$, $N\ge 144$, $N\ge 482$, and $N\ge 1594$, respectively. We can run the following semidefinite program to test for the existence of a state $\rho\in R_1 \cap R_2$:
\begin{equation}
\begin{split}
 \text{find}\qquad&\rho \\
 \text{such that}\qquad
 &\rho\geq 0, \; \tr(\rho)=1,\\
 &-\epsilon(\delta) \mathbb{1}\le \rho-\hat{\rho}_i \le \epsilon(\delta) \mathbb{1}
 \qquad\text{for } i=1,2.
\end{split}
\end{equation}
If the above program does not find a feasible solution, we conclude that the intersection $R_1\cap R_2$ is either empty or does not contain physical states.

\medskip
\paragraph{Intersection for Confidence Region~C$_2$.}
For Confidence Region~C$_2$ we proceed analogously, where here the region with confidence level $1-\delta$ is given by all states $\rho$ satisfying
\begin{equation}
 \norm2{\vec p(\rho)-\vec p(\hat\rho)}^2 \le\epsilon^2(\delta),
\end{equation}
with $\epsilon(\delta)^2 = P_{d^2-1}^{-1}(1-\delta)/(2n)$ and $n$ the number of samples per measurement setting, hence $N=3^qn$. A state $\rho$ in the intersection can be found by the quadratic convex program
\begin{equation}
\begin{split}
\text{find}\qquad&\rho \\
\text{such that} \qquad
 &\rho\geq 0, \; \tr(\rho)=1,\\
 &\sum_{j=1}^{3^q(2^q-1)}\left(\tr(E_j\rho)-\tr(E_j\hat\rho_i)\right)^2\le \epsilon_i^2(\delta)
 \qquad\text{for } i=1,2.
\end{split}
\end{equation}
\medskip
\paragraph{Intersection of Confidence Region~B.}
For regions of type B, we need to find a state in the intersection $\Gamma^{(1)}\cap \Gamma^{(2)}$ of the two confidence polytopes.
The facets of either polytope $\Gamma^{(i)}$, $i=1,2$, are given by the linear inequalities
\begin{equation}
 \tr(E_j\rho)\le f_j^{(i)}+\epsilon^{(i)}_j(\delta),
 \qquad\text{for } j=1,\dotsc 6^q,
\end{equation}
where $\epsilon_j^{(i)}$ is determined by the root of the equation
\begin{equation}
D(f_j^{(i)}\Vert f_j^{(i)}+\epsilon_j^{(i)} )=-\frac1N\log (\delta_j^{(i)}),
\end{equation}
$N$ is the total number of samples, and we make the uniform choice $\delta_j^{(i)}=6^{-q}\delta$.
Finding a state $\rho\in \Gamma^{(1)}\cap \Gamma^{(2)}$ hence corresponds to the semidefinite program
\begin{equation}
\begin{split}
\text{find}\qquad&\rho \\
\text{such that}\qquad
 & \rho\ge 0, \; \tr(\rho)=1,\\
 &\tr(E_j\rho) \leq f_j^{(i)}+\epsilon_j^{(i)}(\delta)
 \qquad \text{for } i=1,2 \text{ and } j=1,\dotsc,6^q.
\end{split}
\end{equation}

\section{Target quantum states}\label{app:states}

The target quantum states used for evaluating the empirical $(1-\delta)$-quantiles of different CRs in Table~\ref{tab:quantiles} are uniformly random pure states sampled according to the Haar measure.
For the tests of intersecting CRs in Figs.~\ref{fig:1and2qubits} and \ref{fig:3and4qubits}, we choose pairs of target states $\rho^{(1)}$, $\rho^{(2)}$. The first state is always chosen as $\rho^{(1)} = \ketbrad{\Psi_{\rm target}}$, where $\ket{\Psi_{\rm target}}$ is obtained by applying uniformly random rigid local unitaries on exemplary pure states of up to 4 qubits, i.e.,
\begin{equation}
    \ket{\Psi_{\rm target}} = U^{\otimes q} \ket{\psi_q} \,,
\end{equation}
where $q=1,2,3,4$ is the number of qubits of $\ket{\psi_q}$, $U\sim \mu_{\rm Haar}(SU(2))$, and
\begin{align}
    \ket{\psi_1} &= \frac{1}{\sqrt{2}} \left(\ket{0}+\ket{1}\right) \,,\\
    \ket{\psi_2} &= \ket{\Phi^+} = \frac{1}{\sqrt{2}} \left(\ket{00}+\ket{11}\right) \,,\\
    \ket{\psi_3} &= \ket{{\rm GHZ}} = \frac{1}{\sqrt{2}} \left(\ket{000}+\ket{111}\right) \,,\\
    \ket{\psi_4} &= \frac{1}{2} \left( \ket{0000}+\ket{0011}+\ket{1100}-\ket{1111} \right) \,.
\end{align}
Then, we distinguish two cases: the second state is either the maximally mixed state $\rho^{(2)} = 2^{-q}\,\mathbb{1}$, or another pure state defined as $\rho^{(2)} = U_z(\theta)^{\otimes q} \rho^{(1)} (U_z(\theta)^{\otimes q})^\dagger$, where $U_z(\theta) = \exp{(i \theta \sigma_z)}$, and $\theta$ is chosen such that $\tr (\rho^{(1)}\rho^{(2)}) = 0.63$.

\new{
\section{Confidence regions using SIC-POVM measurements}\label{app:SICpovm}

Here we analyze confidence regions for tomography schemes based on local SIC-POVMs measurements, instead of Pauli-basis measurements. The single-qubit SIC-POVM is a generalized measurement with effects $E_j=\frac 12\ketbra{\Psi_j}{\Psi_j}$ and
\begin{align}
    \ket{\Psi_1}&=\ket{0} ,\\
    \ket{\Psi_2}&=\frac{1}{\sqrt{3}}\ket{0}+\sqrt{\frac{2}{3}}\ket{1} ,\\
    \ket{\Psi_3}&=\frac{1}{\sqrt{3}}\ket{0}+\sqrt{\frac{2}{3}}e^{\frac{i2\pi}{3}}\ket{1} ,\\
    \ket{\Psi_4}&=\frac{1}{\sqrt{3}}\ket{0}+\sqrt{\frac{2}{3}}e^{\frac{i4\pi}{3}}\ket{1}-
\end{align}
For more than one qubit we use local SIC-POVMs, that is, a single measurement with effects
\begin{equation}
    E_{i_1,i_2,\dotsc}= E_{i_1}\otimes E_{i_2} \otimes \dotsm,
\end{equation}
where $i_k\in \set{1,2,3,4}$.

\begin{figure}
\centering
\includegraphics[scale=0.45]{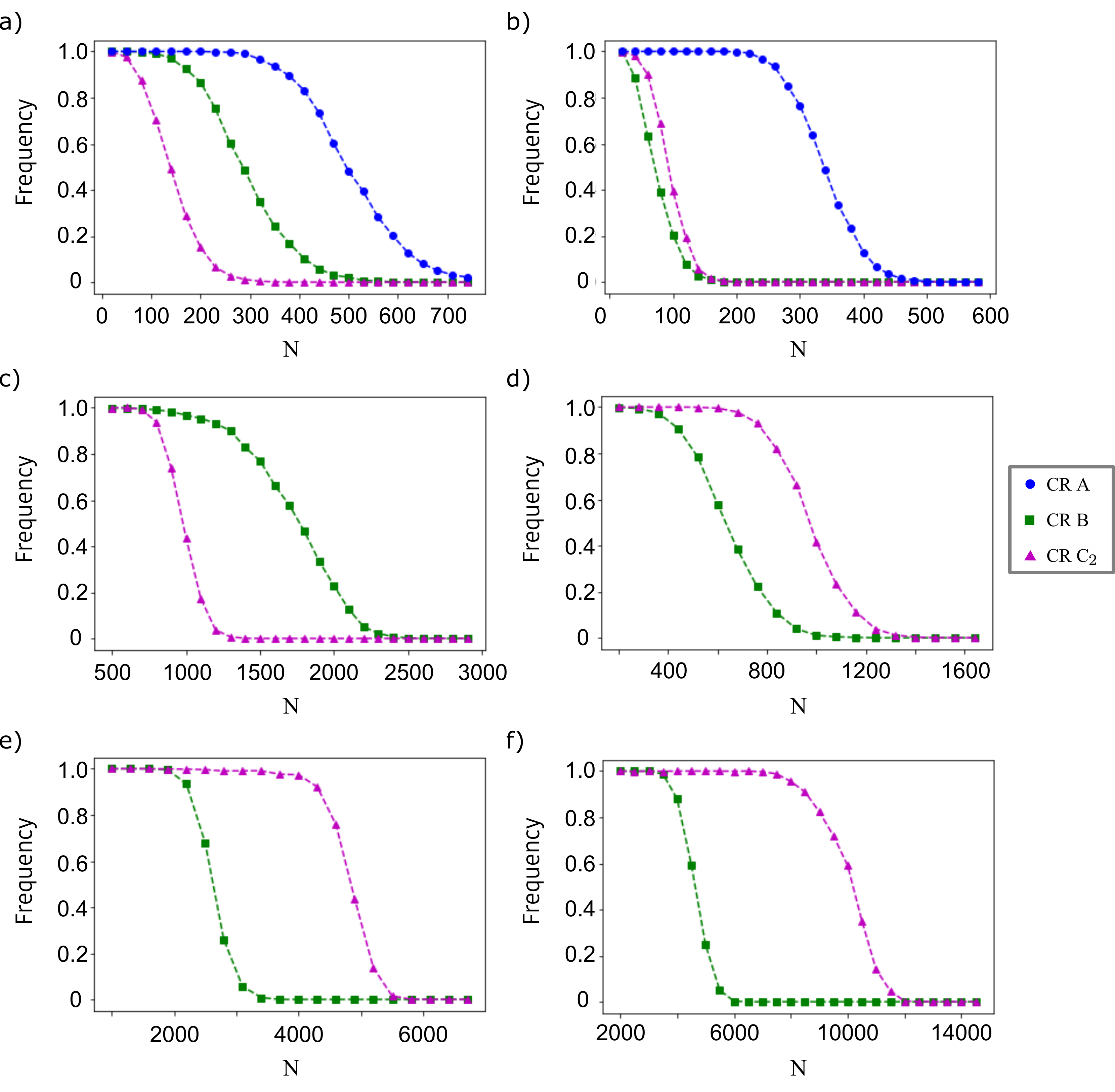}
\caption{\new{Relative frequency of finding a state in the intersection between the CRs using local SIC-POVM measurements of}
a) (1 qubit) a pure state and the completely mixed state,
b) (1 qubit) two pure states with fidelity 0.63,
c) (2 qubits) a rotated Bell state and the completely mixed state, and
d) (2 qubits) two rotated Bell states with fidelity 0.63.
e) (3 qubits) a rotated Greenberger--Horne--Zeilinger (GHZ) state  and the completely mixed state,
f) (3 qubits) two rotated GHZ states with fidelity 0.63.
See also Appendix~\ref{app:states} for details.}
\label{fig:1and2qubitsSIC}
\end{figure}

We show the intersection results of Confidence Region~B, and of Confidence Region~C$_2$ for one and two qubits in Figure~\ref{fig:1and2qubitsSIC}, and of Confidence Region~A for one qubit.
Our analysis is restricted to these cases, as it is not known how to build Confidence Region~A for local SIC-POVMs for more than one qubit. The numerical results for $4$ qubits with 50\% intersection frequency occur at $N$ state preparations as follows: The intersection between a rotated entangled state and the completely
mixed state yields for CR~B $N\sim 13\,000$ and for CR~C$_2$ $N\sim 29\,000$, while the intersection between two rotated entangled states with fidelity 0.63 yield for CR~B $N\sim 13\,000$ and for CR~C$_2$ $N\sim 49\,000$.
\begin{table}
\centering
\begin{tabular}{l|r|r|r|r}
$(1-\delta)$-quantile&1 qubit&2 qubits&3 qubits&4 qubits \\\hline
CR A  &  39\% $\pm$ 2\% &  &   &  \\
CR C$_2$  &  72\% $\pm$ 3\% &  53\% $\pm$ 2\% &  37\% $\pm$ 1\% &  25\% $\pm$ 1\%\\
\end{tabular}
\caption{$(1-\delta)$-quantile of the distribution of the ratio $r$ defined in Eq.~\eqref{eq:quantile} for different CRs and pure Haar-random states for tomography based on SIC-POVM measurements. We used $N=60\,000$ samples on each tomography simulation, and evaluated the $(1-\delta)$-quantile from 10$\,$000 samples of $r$.}
\label{tab:quantilesSIC}
\end{table}
}

\new{
We also perform the $(1-\delta)$-quantile test with local SIC-POVMs on CR C$_2$ in all cases, and on CR A only for one qubit, see Table~\ref{tab:quantilesSIC}. We observe that, although C$_2$ for one qubit is slightly tighter than with local Pauli measurements, it becomes worse as the number of qubits increase. The test for CR A for one qubit is slightly improving.}

\end{document}